# True Bound States in the Continuum in Compact All-Dielectric Structures

Yiyun Yan[1, #], Yichao Liu[1, #, *], Fei Sun[1, †], and Yuxin Zhou[1]

[1] *Key Lab of Advanced Transducers and Intelligent Control System, Ministry of Education and Shanxi Province, College of Physics and Optoelectronics, Taiyuan University of Technology, Taiyuan, 030024 China*

[*] liuyichao@tyut.edu.cn;

[†] sunfei@tyut.edu.cn

[#] These authors contribute equally to this work

**Bound states in the continuum (BICs), known for their theoretically infinite quality ($Q$) factors and strong field localization, hold great promise for high-performance photonic devices. However, conventional true BICs typically rely on infinitely periodic structures, and their realization in finite-sized compact structures faces fundamental challenges, which severely limits device miniaturization and integration. In this work, a compact BIC design method based on optical conformal mapping is proposed, where a conventionally infinite periodic structure extended along one direction is mapped into a finite-sized annular structure. This symmetry transition, i.e., from translational to rotational, enables structural miniaturization while fully preserving the eigenvalues and BIC type of the original system. These transformations require only the adjustment of background permittivity and source distribution, without introducing extreme material parameters. As a concrete example, we show through theoretical and numerical analysis that a transformed compact all-dielectric structure, consisting of a double annular dielectric grating embedded in a gradient-index dielectric background, can support true BICs in a finite region. This work provides a simple and general strategy for achieving true BICs in compact all-dielectric structures, paving the way toward miniaturized high-$Q$ photonic devices.**

Bound states in the continuum (BICs) have attracted extensive attention due to their distinctive physical mechanisms and potential applications [1-3]. Originally proposed by von Neumann and Wigner in quantum mechanics [4], BICs have been extended to classical wave systems, including electromagnetic [5,6], acoustic [7,8], and elastic waves [9]. BICs are nonradiative eigenmodes embedded in the radiation continuum, characterized by a complete suppression of radiative loss, leading to infinite radiative quality ($Q$) factors and vanishing resonance linewidths [1-3]. The high-$Q$ nature enables extreme energy localization and field enhancement. Owing to these properties, BICs have been widely applied in photonic devices, including lasers [10,11], sensors [6,12], and nonlinear optics [13,14].

According to their distinct mechanisms of decoupling from the far-field radiation, BICs are classified into symmetry-protected BICs [15,16] and accidental BICs [17,18]. Symmetry-protected BICs arise from symmetry incompatibility between the eigenmode and the far-field radiation modes. Accidental BICs originate from complete destructive interference of two or more radiative modes in the far field, achieved by tuning system parameters. True BICs typically occur in infinite or periodic structures in at least one direction [1-3], where the number of radiation channels is finite. In such structures, true BICs can be achieved by tuning a finite number of parameters exceeding the number of radiation channels, but this limits device miniaturization and integration. However, true BICs are generally forbidden in compact structures for single-particle-like systems by the non-existence theorem [1]. Such structures have an infinite number of radiation channels, which means that suppressing all far-field radiation by tuning only a finite set of parameters is theoretically impossible. Consequently, true BICs in finite-size structures typically manifest as quasi-BICs (QBICs) with finite $Q$ factors [10-14,19,20]. The only exception is using materials with near-zero or infinite permittivity [21-24]. From the perspective of transformation optics (TO) [25-27], these materials can be regarded as infinitely extending space [28-32] or collapsing it [33,34] along a spatial dimension, thereby circumventing the limitations imposed by the non-existence theorem. However, the intrinsic losses of near-zero-index or metallic materials lead to a drastic reduction in the $Q$ factors [35,36]. In contrast, dielectric materials exhibit intrinsically low losses, making them highly attractive for high-$Q$ photonic devices. Single dielectric resonators are a representative platform for realizing high-$Q$ resonances in compact dielectric structures [19,20], where incomplete destructive interference between the radiation modes of two nearly orthogonal eigenmodes leads to QBICs rather than true BICs. Therefore, at present, realizing true BICs with low-loss compact all-dielectric structures remains challenging.

In this Letter, we adopt the optical conformal mapping [26,35] based on TO, mapping an all-dielectric structure with infinite periodicity along a single direction (supporting true BICs) in the virtual space into a finite-sized, compact all-dielectric structure with angular periodicity in the physical space. The transformation modifies only the system's representation: it converts a structure with translational symmetry into one with rotational symmetry (geometry), a uniform-index into a gradient-index background (material), and a plane-wave into a vortex-beam excitation (source). This transformation alters the field distributions (modes) while preserving the system's eigenvalues, ensuring that the true BICs of the original infinite structure are retained in the new compact structure. This is verified both theoretically and numerically, demonstrating true BICs in a low-loss, compact all-dielectric design.

As an example, we choose a two-dimensional double grating composed of two dielectric gratings with identical periods, embedded in an air background, as the initial BIC-supporting configuration in the virtual space (complex coordinate $\eta = u+iv$ in Fig. 1(a)). The double grating is oriented along the $v$-direction, with structural period $D$, inter-grating distance $d$, dielectric unit dimensions $d_1$ ($u$-direction) and $h$ ($v$-direction), and refractive index $n$. A TE-polarized plane wave is incident on the double grating at an angle $\theta$, with the electric field polarized along the $z$-direction. True BICs can be formed for specific incident angles, structural parameters, and material parameters [5]. To transform this virtual structure into physical space (complex coordinate $\zeta = x+iy$), we apply the following coordinate transformation:

$$\zeta = e^{a\eta}, \quad (1)$$

where $a = 2\pi/(mD) = k_v/l$, $m$ is the number of unit cells in the annular grating ($m$ is a positive integer), $k_v$ is the component of the incident plane wave vector along the $v$-direction in the virtual space, and $l$ denotes the topological charge of the vortex beam in the physical space. In the virtual space, the strip domain with $u \in [u_1, u_2]$ and $v \in (-\infty, +\infty)$, containing the double grating, serves as the core transformation region. Under the coordinate transformation in Eq. (1), the straight boundaries $u = u_1$ and $u = u_2$ are mapped onto the circular boundaries $C_1$ and $C_2$, respectively. Here, $u_2 = -\ln(a)/a$ is chosen to ensure a unit refractive index at $C_2$ in the physical space, achieving impedance matching with the surrounding air and thereby minimizing reflections. $u_1 = u_2 - L$, where $L$ is chosen to be sufficiently large to accommodate the near-field region in which the electromagnetic modes (bound states) of interest exist.

A double grating with fixed parameters is chosen as a representative case in this work. The underlying mechanism is general and not restricted to this specific choice. After the transformation, the double dielectric grating in the virtual space is mapped into the double annular dielectric grating in the physical space, and the air background is converted into the gradient-index background, resulting in a finite-sized compact all-dielectric structure [Fig. 1(b)]. The corresponding refractive index distribution in the physical space is presented in Fig. 1(c), which is given by (see the Supplementary Materials (SM) [37]):

$$n(\zeta) = n(\eta) \frac{1}{|a\zeta|}. \quad (2)$$

Note that through the transformation in Eq. (1), a plane wave that originally excites the double grating at $u = u_1$ in the virtual space is transformed into a vortex beam [42] with the topological charge $l = k_v/a$ located on $C_1$ in the physical space, as indicated by red arrows in Figs. 1(a) and 1(b). In addition, the $E_z$ distribution in the virtual space is also correspondingly transformed into that in the physical space via the coordinate transformation. Therefore, the relationship between the electric field distributions in the two spaces can be established through TO, i.e., $E_z(\zeta) = E_z(\eta)$ (see SM [37] for detailed derivation).

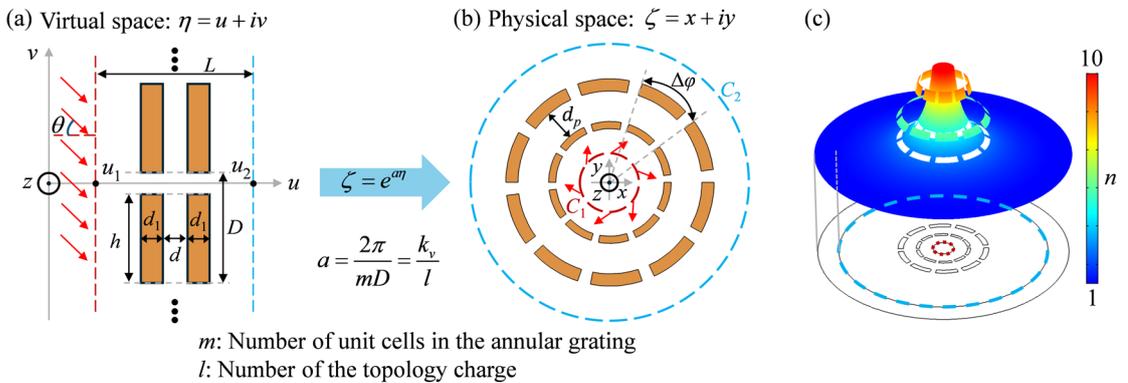

FIG. 1. Schematic of the studied structure and parameter distributions. (a) Double grating in air

background in the virtual space, with $h = 0.8D$, $d_1 = 0.2D$, dielectric refractive index $n = 2$, $m = 10$, and $L = 4D$. (b) Compact all-dielectric structure, consisting of a double annular grating embedded in a gradient-index background in the physical space. (c) Refractive index distribution in the physical space.

Next, we verify the existence of BICs via numerical simulations, starting with the case of $l = 0$, where a plane wave is normally incident on the double grating in the virtual space. After the transformation to the physical space, the grating geometry is modified, leading to corresponding changes in both the inter-grating distance and the period. Specifically, in the physical space, the inter-grating distance is given by $d_p = (e^{ad/2} - e^{-ad/2})/(ae^{aL/2})$ and the spatial period $D$ transforms into the angular period $\Delta\varphi = 2\pi/m = aD$, as derived from Eq. (1). For direct comparison and simplicity, all figures and theoretical analyses in this work use the parameters in the virtual space. Fig. 2(a) shows the transmittance $T$ in the virtual space (left) and the physical space (right) as a function of the normalized total wave vector $k/G$ and normalized distance $d/D$, with the reciprocal lattice vector $G = 2\pi/D$ in the virtual space. The two panels of Fig. 2(a) are nearly identical, as TO performs a coordinate mapping that leaves the eigenvalues of the system invariant while transforming the spatial profiles of the eigenfunctions, a true BIC condition defined by a real eigenfrequency within the radiation continuum is preserved under the transformation. Vanishing resonance linewidths are observed at $d = 0.41D$ and $d = 1.02D$, indicating signatures of true BICs (highlighted by blue circles and labeled as modes I–IV). As $d$ decreases, two distinct branches exhibit the avoided crossing behavior.

The formation mechanism of BICs is theoretically analyzed and numerically verified by plotting the transmittance of the single and double annular gratings versus $k/G$ at $d = 0.41D$ and $d = 1.02D$ in the physical space. An annular dielectric slab supports waveguided modes with infinite lifetimes via total internal reflection, which are discrete bound states outside the radiation continuum. Patterning the slab into an annular grating provides momentum compensation through the reciprocal lattice vector $G_\theta$. This allows waveguide modes to match the momentum in free space, $k_\theta = k_{wg} - pG_\theta$, thereby forming resonances that can couple to radiation modes in free space. Here, $p$ is the diffraction order, $k_\theta$ is the angular wave vector component in free space, and $k_{wg}$ is the propagation constant of the annular waveguide modes. $G_\theta = 2\pi/(R\Delta\varphi) = G/(aR)$, and $R$ is the effective radius of the annular grating. The discrete rotational symmetry of the annular grating quantizes the far-field radiation channels into an infinite set of discrete diffraction channels, each carrying a specific quasi-angular momentum. For subwavelength gratings, only the zeroth-order diffraction channel is open, while other higher-order diffraction channels are closed (evanescent waves do not radiate). Consequently, a single annular grating supports a single Fano resonance that couples to the zeroth-order diffraction channel and radiates outward [Figs. 2(b) and 2(c), left panel]. When a second annular grating (transformed from an identical grating in the virtual space to ensure identical resonances of the two annular gratings) is introduced, their resonances hybridize through the same open channel. By tuning a limited set of parameters, the radiation from the two hybridized eigenmodes can perfectly destructively interfere in the far field. In the double annular grating [Figs. 2(b) and 2(c), right panel], only a single Fano resonance with an approximately doubled linewidth is observed. This phenomenon arises from the hybridization of two identical single-grating resonances. The disappearance of one hybridized mode's resonance signifies the formation of a true BIC—a dark mode with vanishing linewidth, as marked by black dashed lines, while the other evolves into a superradiant mode with a doubled linewidth.

It is worth noting that the BICs supported by this system exhibit distinct field distributions. Figs. 2(d) and 2(e) show the $E_z$ field distributions of the eigenstates corresponding to modes I–IV, which in the physical space closely correspond to those in the virtual space. Modes I and III display symmetric profiles with in-phase fields in the two gratings, while modes II and IV exhibit antisymmetric profiles with out-of-phase fields. Although modes I and II in the virtual space are true BICs at the Γ point ($k_v = 0$), their formation mechanism is not symmetry-protected. This is because they can couple to radiation channels of the same parity: modes I and II exhibit even symmetry along the $v$-direction, while modes III and IV are even along the angular direction. In both cases, the modes match the parity of the radiation fields, allowing them to radiate outward. Their complete decoupling from radiation instead arises from perfect destructive interference of the far-field waves, achieved by tuning parameters. Therefore, these BICs are classified as accidental BICs. Specifically, they belong to the Fabry–Pérot BICs (FP-BICs), where two identical resonances are spatially separated and true BICs are achieved by tuning the distance between them [1]. Notably, the FP-BIC character is preserved in both the virtual and physical spaces, demonstrating that the BIC type remains unchanged via TO.

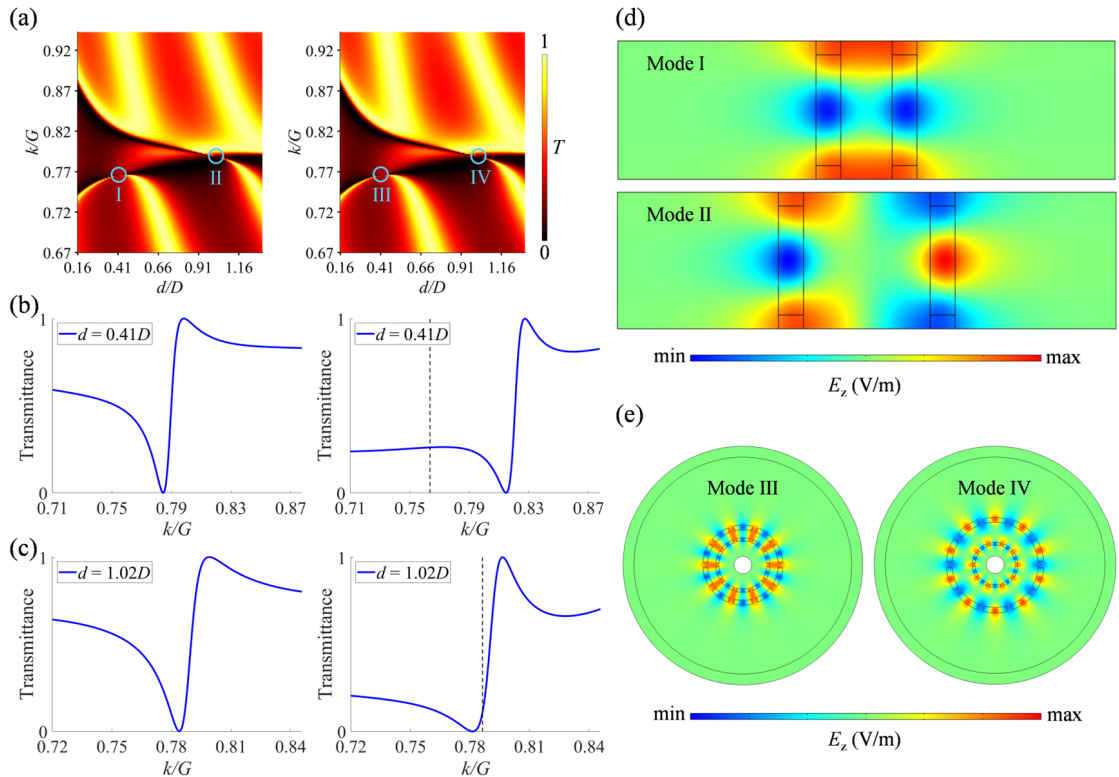

FIG. 2. Formation of BICs for $l = 0$. (a) The transmittance $T$ as a function of the normalized total

wave vector $k/G$ of the incident source and normalized distance $d/D$ under normal incidence ($k_v = 0$) in the virtual space (left) and for $l = 0$ in the physical space (right). The BICs are marked by blue circles. (b) and (c) The transmittance versus $k/G$ in the physical space at (b) $d = 0.41D$ and (c) $d = 1.02D$. The left and right panels correspond to the single and double annular gratings embedded in a gradient-index background, respectively. (d) and (e) $E_z$ field distributions of the eigenstates corresponding to modes I, II, III and IV.

For nonzero $l$, a plane wave is obliquely incident on the double grating in the virtual space with a fixed $k_v$. As shown in Figs. 3(a) and 3(b), both the avoided crossing behavior and the vanishing resonance linewidth (FP-BIC, highlighted by blue circles) are observed. A broad resonance is clearly visible above each BIC. The FP-BICs can be explained using the temporal coupled mode theory (TCMT) (see SM [37]).

To further verify the existence of BICs, we calculate the transmittance of the system for different $d$ with fixed $l = 1$ and $l = 2$, as shown in Figs. 3(c) and 3(d). At $d = 0.53D$ and $d = 0.64D$, no resonances are observed. As $d$ deviates from these values, Fano resonances reappear as QBICs whose linewidths broaden with increasing deviation.

True BICs cannot be directly excited, but slight perturbations can transform them into excitable QBICs. Their extremely high $Q$ factors arise from strong but imperfect destructive interference, enabling both field localization and enhancement despite leakage. To illustrate these features, Figs. 3(e) and 3(f) show $E_z$ field and Poynting vector distributions for ($l = 1$, $d = 0.51D$, $k = 0.67G$) and ($l = 2$, $d = 0.62D$, $k = 0.61G$), respectively. For a unit-amplitude incident vortex beam, the electric field is strongly enhanced and localized within the gratings. Notably, the number of angular periods of the field distributions matches that of the incident source, characterized by $l$.

In the physical space, the Poynting vector of the incident vortex beam contains both transverse and radial components [Fig. S1 in SM [37]]. To achieve true BICs within the annular region, the radial energy flow must be completely suppressed. As shown in Figs. 3(e) and 3(f), the double annular grating drastically reshapes the energy flow of the incident vortex beam: the originally dominant outward radial component is strongly suppressed, and the energy is redirected into a transverse (azimuthal) component localized within the grating region. Since the transverse component does not couple to the radiative radial channel, the electromagnetic wave is effectively localized within the double annular grating.

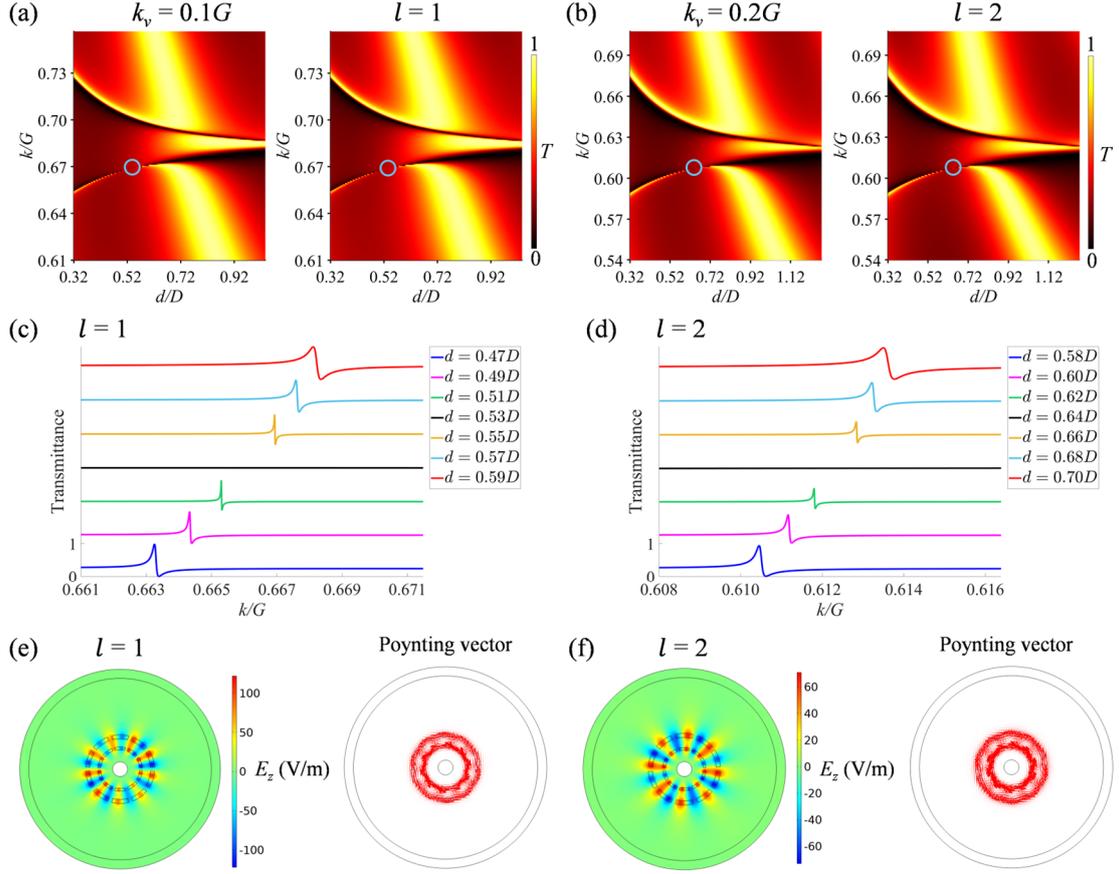

FIG. 3. Formation of BICs for $l = 1$ and $l = 2$. (a) and (b) The transmittance $T$ versus $k/G$ and $d/D$ under oblique incidence in the virtual space with (a) $k_v = 0.1G$ and (b) $k_v = 0.2G$ (left). The corresponding results in the physical space excited by vortex beams with (a) $l = 1$ and (b) $l = 2$ (right). BICs are indicated by blue circles. (c) and (d) The transmittance in the physical space versus $k/G$ for vortex beams with (c) $l = 1$ and (d) $l = 2$ at different $d$. (e) and (f) $E_z$ field and corresponding Poynting vector distributions of QBICs for (e) $l = 1$ and (f) $l = 2$, demonstrating strong field localization and enhancement. The red arrows represent the Poynting vectors.

Can true BICs be achieved in an air background by introducing a double annular grating structure? To investigate this, we analyze a simplified structure where only the high gradient-index regions of the double annular grating (orange regions in Fig. 1(b)) are retained, while other gradient-index regions (white regions in Fig. 1(b)) are replaced with air. Comparing Fig. 4(a) and Fig. 4(b) shows that, within the same parameter space, BICs exist in the TO-designed structure [Fig. 4(a)] but disappear in the simplified structure [Fig. 4(b)]. From the perspective of TO, achieving an air background in the physical space requires the gradient-index background in the virtual space (varying along the $u$-direction). This gradient variation causes the refractive index around each grating to change differently, resulting in non-identical resonances for the two gratings. As a result, complete destructive interference in the far field is prevented, suppressing the formation of a true BIC. In the simplified structure, the resonances vary with $d$ (as $d$ varies with $u$), making it extremely difficult to realize a true BIC with two annular gratings in an air background. This is because any modification destroys the true BIC condition in parameter space, making it harder to access or even eliminating it entirely. In contrast, TO-designed structures are easy to find true BICs because the resonances of the two gratings remain identical regardless of $d$. These results further highlight the critical role of the TO-designed gradient-index background. In theory, although true BICs may exist for the double annular grating in an air background, locating them in parameter space is highly nontrivial. TO instead provides a convenient method to achieve true BICs in compact structures. This approach extends true BICs in periodic structures such as gratings, offering a practical route for designing compact true BICs. Furthermore, compact true BICs can be transformed into dynamically tunable and more practical QBICs through active refractive-index tuning, such as time-varying modulation [43], providing a promising route toward tunable photonic applications (a detailed discussion is provided in SM [37]).

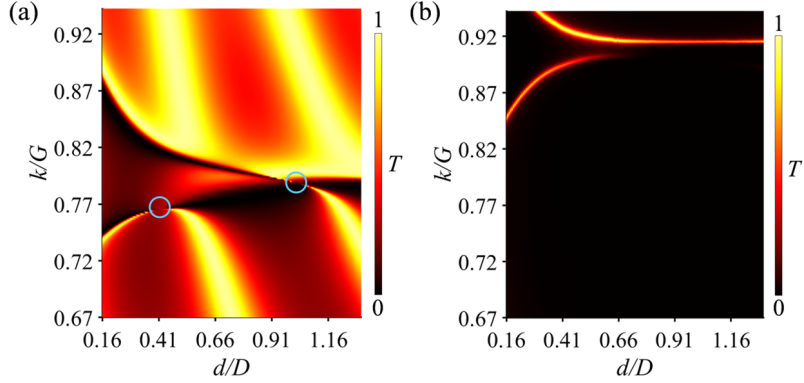

FIG. 4. The transmittance $T$ versus $k/G$ and $d/D$ for $l = 0$ at different structures. (a) TO-designed structure with the gradient-index background. The BICs are marked by blue circles. (b) The simplified structure in which the gradient-index background is replaced by air.

In conclusion, we establish a theoretical framework to convert infinite all-dielectric structures supporting true BICs into compact all-dielectric structures via TO. The existence of true BICs is confirmed by both analytical calculations and numerical simulations in the compact all-dielectric structure (transformed from a one-dimensional infinite periodic double dielectric grating in an air background), which consists of a double annular dielectric grating embedded in a properly designed local gradient-index dielectric background. This transformation, which converts translational symmetry to rotational symmetry, preserves the eigenvalues and BIC type of the original system by only adjusting the background permittivity (from uniform to radially varying) and source distribution (from plane waves to vortex beams), without relying on extreme material parameters. Our findings demonstrate a simple and general strategy for achieving true BICs in low-loss, compact all-dielectric structures, laying the foundation for the development of miniaturized high-$Q$ photonic devices.

The authors acknowledge the financial support by the National Natural Science Foundation of China (Nos. 12374277, and 12274317), San Jin Talent Support Program—Shanxi Provincial Youth Top-notch Talent Project, the Natural Science

Foundation of Shanxi Province (202303021211054), and Shanxi Province Higher Education Institutions Young Faculty Research and Innovation Support Program (2025Q006).

The authors declare no competing financial interests.

The main data and models supporting the findings of this study are available within the paper. Further information is available from the corresponding authors upon reasonable request.